\documentclass[prl,twocolumn,letterpaper,10pt,twoside,tightenlines,nofootinbib,showpacs]{revtex4}
\usepackage{amsmath,amsfonts,amssymb,latexsym}
\usepackage{graphicx,xcolor}
\usepackage[sort&compress]{natbib}

\begin{document}

\newcommand{\eg}{{\it e.g.}}
\newcommand{\cf}{{\it cf.}}
\newcommand{\etal}{{\it et. al.}}
\newcommand{\ie}{{\it i.e.}}
\newcommand{\be}{\begin{equation}}
\newcommand{\ee}{\end{equation}}
\newcommand{\bea}{\begin{eqnarray}}
\newcommand{\eea}{\end{eqnarray}}
\newcommand{\bef}{\begin{figure}}
\newcommand{\eef}{\end{figure}}
\newcommand{\bce}{\begin{center}}
\newcommand{\ece}{\end{center}}
\newcommand{\red}[1]{\textcolor{red}{#1}}
\newcommand{\RR}[1]{\textcolor{purple}{\bf #1}}
\newcommand{\ZT}[1]{\textcolor{red}{#1}}
\newcommand{\dd}{\text{d}}
\newcommand{\ii}{\text{i}}
\newcommand{\lsim}{\lesssim}
\newcommand{\gsim}{\gtrsim}
\newcommand{\RAA}{R_{\rm AA}}
\newcommand{\QQb}{Q\bar{Q}}
\newcommand{\ttb}{t\bar{t}}
\newcommand{\Zpol}{Z_{\rm pole}}

\title{Toponium Spectrum in the Complex-Energy Plane}

\author{Zhanduo~Tang$^{1,2,3}$, Oliver Fast$^{1,4}$, and Ralf~Rapp$^1$}
\affiliation{$^1$Cyclotron Institute and Department of Physics and
Astronomy, Texas A\&M University, College Station, TX 77843-3366, U.S.A. 
\\
$^2$Key Laboratory of Nuclear Physics and Ion-beam Application (MOE), Institute of Modern Physics, Fudan University, Shanghai 200433, China 
\\
$^3$Shanghai Research Center for Theoretical Nuclear Physics,
NSFC and Fudan University, Shanghai 200438, China
\\
$^4$Department of Physics, University of North Texas, Dallas, TX , U.S.A.
}

\date{\today}

\begin{abstract}
{The electroweak decay width of the top quark of $\sim$2\,GeV is comparable to binding energies expected for the toponium system, raising the long-standing question of whether top and antitop quarks can sustain meaningful bound states. This issue has been been reignited by recent discoveries of a cross section enhancement near the $t\bar t$ threshold by the CMS and ATLAS collaborations in $pp$ collisions at the LHC.  Here, we deploy a $T$-matrix approach with an underlying Cornell potential to study the properties of the toponium system by varying the top-quark widths in the intermediate propagators. In particular,  we carry out a complex-energy analysis of the $T$-matrix to assess how its poles, as a rigorous criterion for bound-state formation, develop from the limit of small widths to realistic ones. We also revisit the impact of the confining force in the potential on the toponium $T$-matrix.        
}
\end{abstract}


\maketitle

{\it Introduction.---}
In Quantum Chromodynamics (QCD) the top quark ($t$) occupies a special role. Its extremely large mass, $m_t\simeq 173$\,GeV, implies that its electroweak decay has a large phase space and thus generates a width, $\Gamma_t\simeq 1.4$\,GeV~\cite{Strassler:1990nw,Kumar:2024xsv,Beneke:2024sfa,Yan:2024hbz,ParticleDataGroup:2024cfk,Bai:2025buy}, which turns out to be comparable to the typical hadronic scale of 1\,GeV, and significantly larger than most strong-decay widths of hadronic resonances (which are of order 1/fm = 0.2\,GeV). This leads to the naive expectation that top quarks do not have time to form hadronic bound states~\cite{Bigi:1986jk,Kuhn:1980gw,Beneke:2024sfa}. Even for the possibly strongest bound system, the $\ttb$ toponium, the ground-state binding energy is expected to be around $E_B\simeq$\,2\,GeV and thus smaller than the sum of the widths of the top and anti-top quark driving its decay~\cite{Thompson:2025cgp,Lopez:2025kog}. 

Nevertheless, calculations for $\ttb$ production in collider experiments still predict a large threshold enhancement as a result of the strong final-state interaction, leading to a broad peak structure (sometimes referred to as ``quasi-bound state")  in the observed cross section~\cite{Fadin:1990wx,Strassler:1990nw,Ju:2020otc,Fuks:2024yjj}, see also Ref.~\cite{Fuks:2025sxu} for a recent review. These predictions were recently corroborated by the CMS~\cite{CMS:2025kzt} and ATLAS~\cite{ATLAS:2026dbe} collaborations in pp collisions at the LHC, who reported a cross section enhancement that is more than 5 standard deviations above baseline perturbative QCD (pQCD) production and consistent with the formation of a quasi-bound state in the pseudoscalar toponium channel, $\eta_t$. Clearly, the identification of individual toponium states from this broad single-peak structure and the related issue of whether a ``true" bound state has formed, will be challenging at best~\cite{Bai:2025buy,Pintucci:2026agl,Beneke:2024sfa,Thompson:2025cgp,CMS:2024pts}. 

However, there is a well established method in hadron spectroscopy that is widely regarded as the most rigorous way to identify the presence of bound states, namely a pole analysis of pertinent $T$-matrix amplitudes in the complex energy plane. Usually this applies to situations where coupled channels exist and a bound state in a heavier channel can decay into a lighter channel (\eg, a $K\bar K$ bound state that decays into the $\pi\pi$ channel). The situation is slightly different if the constituents themselves are unstable, as, \eg,  encountered when a bound state is immersed into a medium where the constituents acquire complex selfenergies. This has recently been investigated for the case of quarkonia in the quark-gluon plasma (QGP), to address the long-standing problem of their survival in the presence of large dissociation widths~\cite{Tang:2025ypa}. By analyzing the poles of the in-medium $\QQb$ $T$-matrices in the complex energy plane, it was found that bottomonium bound states can survive rather deep into the QGP, even if their nominal binding energy is relatively small or even vanishing,  in any case much smaller than the collisional widths of the bottom quarks. As this problem shares several features with that of toponium in vacuum, the method of Ref.~\cite{Tang:2025ypa} appears to be well suited to assess the latter's bound-state spectra in vacuum. 

Recent papers studying the toponium spectrum can be found in Refs.~\cite{Jiang:2024fyw,Wang:2024hzd,Zhang:2026iuw,Fuks:2025sxu,Fu:2025yft,Lopez:2025kog,Thompson:2025cgp}.
In the present paper we largely follow the early work where toponia were studied in a Lippmann-Schwinger equation~\cite{Strassler:1990nw}, with a focus on how unstable top quarks manifest themselves in observable cross sections and how the pertinent threshold enhancement can be used as quantitative measure of the top-quark mass and width, as well as the strong-coupling constant in the underlying heavy-quark potential. Here, our main objective is to scrutinize toponium properties with respect to (academic) variations in the top quark width to obtain a better understanding of the underlying bound-state structure. While for small widths, $\Gamma_t\ll E_B$, the bound-state structure can be readily obtained from the imaginary part of the $t\bar t$ $T$-matrix (or the corresponding spectral functions) in the various quantum number channels, this becomes much less obvious, even impossible, if the width becomes comparable to, or larger than, the level splittings. However, by deploying a complex-energy analysis we can identify poles of the $T$-matrix  and thus unambiguously determine the presence of bound states even for large widths. In fact, the real and imaginary part of the complex energy, $Z = E_R + i E_I$, at the pole position directly provide, respectively, the bound-state energy and its total width, $\Gamma_{\ttb} = 2E_I^{\rm pole}$.

In the remainder of this paper, we briefly review the $T$-matrix approach as used in our previous works for charmonia and bottomonia, investigate the toponium spectrum in the small-width limit including an evaluation of the impact of the string force, carry out the complex-pole analysis to assess bound-state formation under realistic conditions, and conclude.

{\it $T$-matrix Approach.--}
The two-body $T$-matrix equation can be derived from the 4-dimensional Bethe-Salpeter equation in the static limit (\ie, for small energy transfer, $q_0\ll q$). After partial-wave expansion one has 
\bea
T^\alpha_{Q\bar Q}(E,p,p') &=& V^\alpha_{Q\bar Q}(p,p') + \frac{2}{\pi} \int k^2 dk V^\alpha_{Q\bar Q}(p,k) \nonumber \\
&&\times G_{Q\bar Q}(E,k)  T^\alpha_{Q\bar Q}(E,k,p'), 
\label{tmat}
\eea
for each quantum number $\alpha$ (total spin, angular momentum, color, isospin) of the two-particle system. In the following we focus on $S$-wave states and neglect hyperfine (spin-spin) splittings. The Lippmann-Schwinger equation above resums the ladder series of an underlying potential
for which we employ the standard Cornell potential (color singlet),
\be
\widetilde{V}(r) = -\frac{4}{3} \frac{\alpha_{s}}{r}  +\sigma r \ .
\ee 
We adopt the set-up from our previous works where $\alpha_s$=0.27 and $\sigma$=0.225\,GeV$^2$ are chosen to reproduce the vacuum free energy from lattice QCD~\cite{HotQCD:2014kol},  with a (smooth) string breaking at $r_s\simeq 1$\,fm~\cite{Liu:2017qah}. After Fourier transform into momentum space, we account for an effective running of $\alpha_s$~\cite{Riek:2010fk}, relativistic corrections from magnetic interactions (which are negligible for toponium) and a small-momentum cutoff to tame the infrared singularity of the Coulomb potential. This setup yields a fair description of the vacuum spectrum of charmonium, bottomonium and $B_c$ hadrons~\cite{Tang:2023lcn,Wu:2023djn}.
The quark-antiquark propagator in Eq.~(\ref{tmat}) is obtained from a folding integral, 
\be
G_{Q\bar Q}(E,k)=\int dk_0 G_{Q}(E-k_0,k)G_{\bar Q}(k_0,k) \ ,  
\label{GQQ}
\ee
with single-quark propagators
\be
G_{Q}(k_0,k) =  1/[k_0-\varepsilon_{Q}(k) + i \Gamma_Q/2 ]  \ .
\label{GQ}
\ee
The width of the top quark is taken as a constant since its energy will be in the vicinity of its mass implying that energy-momentum dependencies are small, but we will vary its value. For definiteness, we use a  (bare) top-quark mass of $m_t^0$=173\,GeV, which is in the range of recent measurements although slightly above the average quoted by the particle data group, 172.60$\pm$0.27\,GeV~\cite{ParticleDataGroup:2024cfk}, We note that the bare mass is augmented by an effective selfenergy contribution induced by the infinite-distance limit of the potential, \ie, $\Delta m_t = V_\infty/2 = 1.1/2$\,GeV = 0.55\,GeV, corresponding to a $\ttb$ threshold of $2m_t\sim$347.1\,GeV. 

{\it Toponium Spectrum.--}
%
\begin{figure}
    \centering
    \includegraphics[width=1\linewidth]{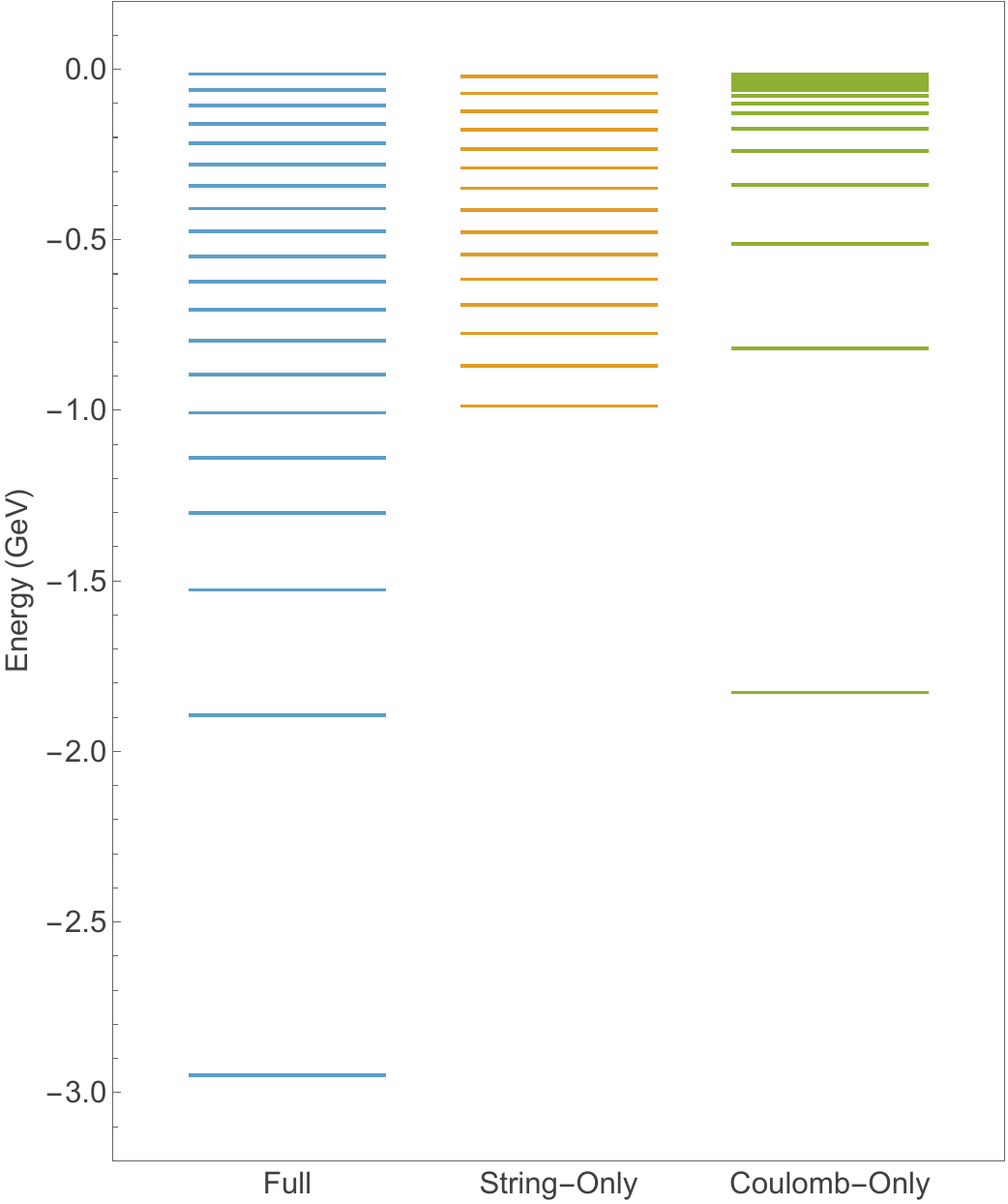}
    \caption{Level spectrum of binding energies for $S$-wave toponium in the small top-quark width limit ($\Gamma_t=$5\,MeV). The binding energies are shown for 3 input interactions: full Cornell potential, string interaction only and  color Coulomb-only, with $\alpha_s=0.27$ and $\sigma=0.225$ GeV$^2$.}
    \label{fig:levels}
\end{figure}
We start our analysis by using an artificially small top quark width of $\Gamma_t=5$\,MeV to properly resolve the ``would-be" bound-state spectrum of the $\ttb$ system. 
The pertinent energy level spectrum is shown in Fig.~\ref{fig:levels} for three variants of the input interaction: the full Cornell potential as well as string or color-Coulomb interaction only. We note that for the latter the effective-mass term, added to the bare top quark mass generated by the finite infinite-distance limit of the string term, is absent.
For the full interaction, the ground-state (GS) binding energy  amounts to $\sim$2.9\,GeV, which is very comparable to the result of 2.8\,GeV from a Schr\"odinger equation where a perturbative color-Coloumb potential at short distance was matched to a Cornell potential at $r>0.1 fm$ and $m_t=$172.4\,GeV~\cite{Bai:2025buy}; it is also comparable to a Schr\"odinger equation evaluation with a Cornell potential resulting in a slightly lager GS binding of 3.5 GeV (presumably due to a larger $\alpha_s$)~\cite{Jiang:2024fyw}.
Our GS binding energy is significantly larger than the 1.8\,GeV found Ref.~\cite{Wang:2024hzd} using an instantaneous Salpeter equation with a Cornel potential, possibly due to differences in $\alpha_s$. In the Dyson-Schwinger approach of Ref.~\cite{Zhang:2026iuw}, which includes a color-Coulomb plus soft interaction that was previously found to describe charmonium and bottomonium spectra, the toponium GS mass was estimated in the range of 344-346\,GeV, where the uncertainty is primarily driven by the top quark mass; this is compatible with our result of 344.2\,GeV.
Our level spectrum features a rather rich set of excited states that initially exhibits decreasing gaps characteristic for a Coulomb-like spectrum but subsequently transits into an essentially constant spacing signaling the prevalence of the confining force all the way to the $\ttb$ threshold. This can be highlighted when switching off the Coulomb term in the potential: the ground state loses about two third of its binding energy while the rather dense set of spectral lines toward threshold persists, as can be clearly seen in the level spectrum shown in Fig.~\ref{fig:levels}. The impact of the (lack of) the string tension is a much stronger effect than found in previous studies. In turn, when switching off the string interaction, the ground state loses about 1/3 of its binding and the gaps in the excited states are now rapidly decreasing toward threshold. 

\begin{figure}
    \centering
    \includegraphics[width=0.99\linewidth]{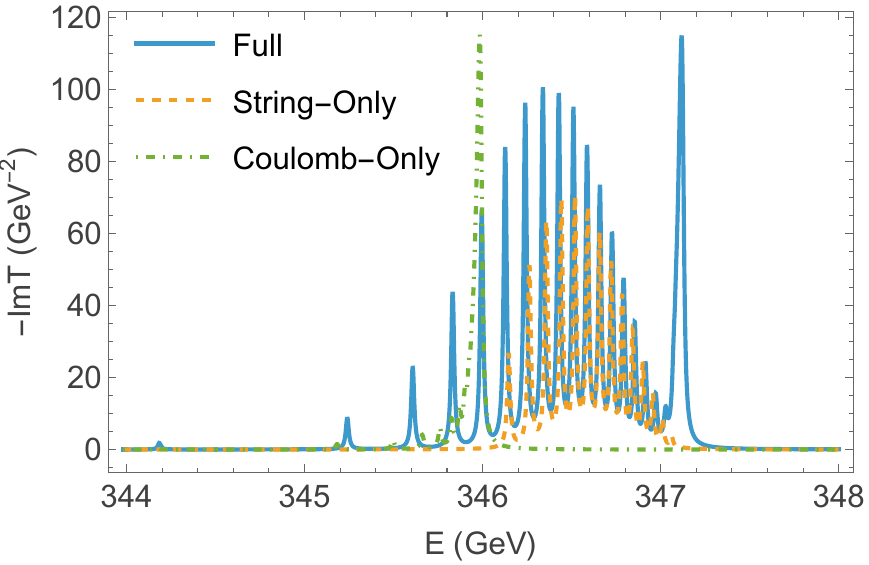}
    \caption{Imaginary part of the $S$-wave toponium $T$-matrix for the three potential scenarios described in the text.}
    \label{fig:ImT-comp}
\end{figure}

The underlying imaginary part of the $T$-matrix in the small width limit is displayed on an absolute scale (\ie, including the pertinent thresholds) in Fig.~\ref{fig:ImT-comp}. Two features are noteworthy that are not apparent from the level spectrum: (a) the full spectrum develops remarkable strength in the string-dominated part of the spectrum while the ground states in either scenario carry comparatively little strength; 
(b) the Coulomb potential develops a rather pronounced near-threshold peak (which is absent in the string-only scenario), presumably due to the dense level spectrum of the $1/r$ potential in this region.

\begin{figure}
    \centering
    \includegraphics[width=0.99\linewidth]{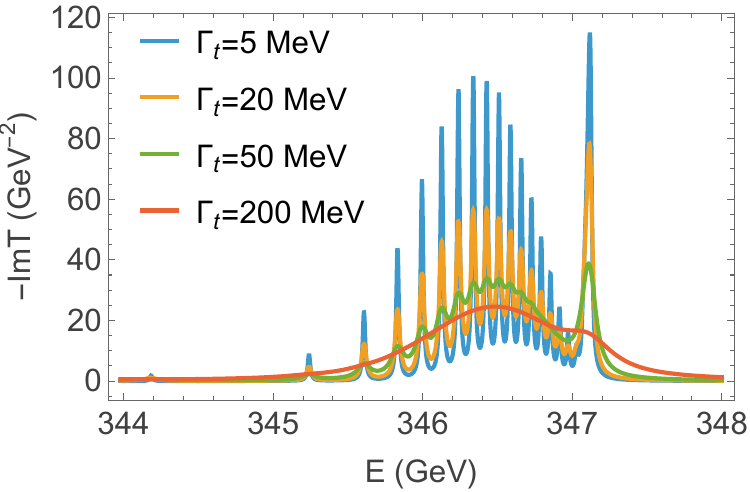}
    \caption{$S$-wave toponium $T$-matrix for variable top-quark widths.}
    \label{fig:ImT-gamma}
\end{figure}
Next, we study the effect of the top-quark width on the spectrum for the full Cornell interaction, cf.~Fig.~\ref{fig:ImT-gamma}.
The $S$-wave bound states remain well defined and readily identifiable at $\Gamma_t$=20\,MeV while at 50\,MeV the overlap of the peaks is already large, with only a slight residual oscillation left on top of a broad maximum structure. At $\Gamma_t$=200\,MeV, the spectrum has ``melted" into a broad maximum structure with no trace of individual bound states. It therefore appears that once the top-quark width surpasses the typical energy spacing of the bound states, the latter's structure information has been lost.

\begin{figure}[t]
    \centering
    \includegraphics[width=0.95\linewidth]{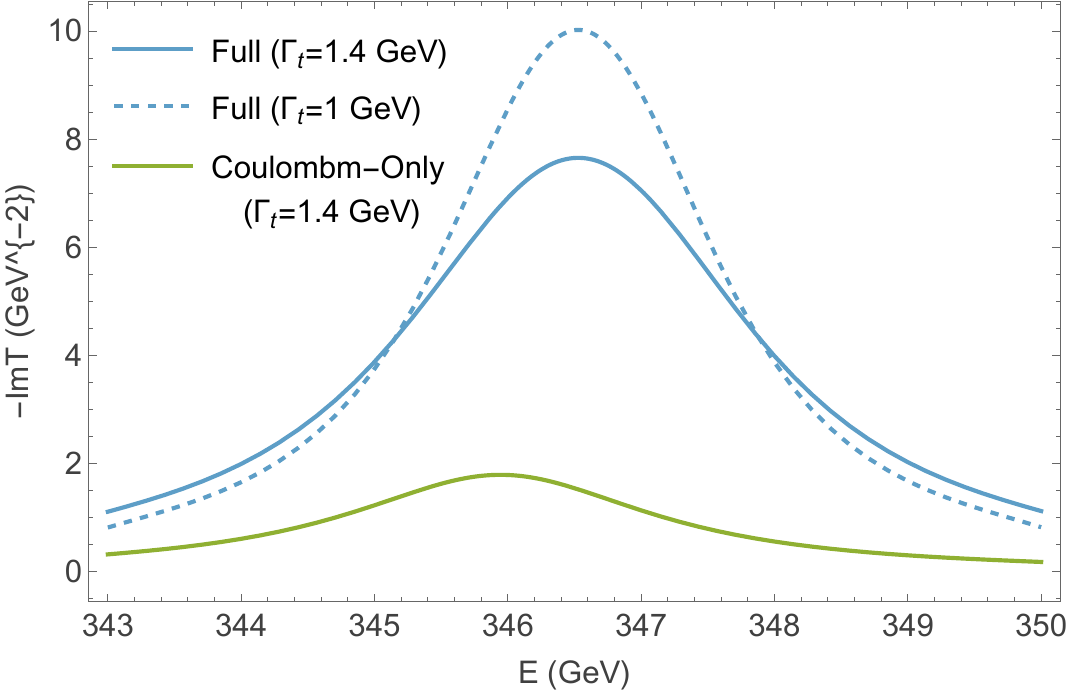}
    \caption{$S$-wave $T$-matrix (imaginary part) in our baseline scenario with a realistic top-quark width of 1.4\,GeV.}
    \label{fig:ImT-14}
\end{figure}
We finally show our results for a realistic top-quark width of $\Gamma_t$=1.4\,GeV, cf.~Fig.~\ref{fig:ImT-14} for the pertinent imaginary part of the $T$-matrix. The spectrum consists of a single broad peak centered around an energy of about 0.5\,GeV below the nominal $\ttb$ threshold of 347.1~GeV, without any discernible substructure including the threshold itself . 
In particular the location of the maximum is about 2\,GeV above the ground-state energy in the small-width limit, and as such is much more indicative of the densely populated spectrum that is driven by the confining force, as well as the threshold enhancement.
Even in the Coulomb-only case, the maximum in the imaginary part of the $T$-matrix is approximately 1\,GeV above the (would-be) bound-state mass.
This raises the question whether there are, in principle, any bound states left. Toward addressing this question we proceed to analyze the $T$-matrix in the complex energy plane following the method introduced in Ref.~\cite{Tang:2025ypa}.

\begin{figure}[t]
    \centering
   \includegraphics[width=0.95\linewidth]{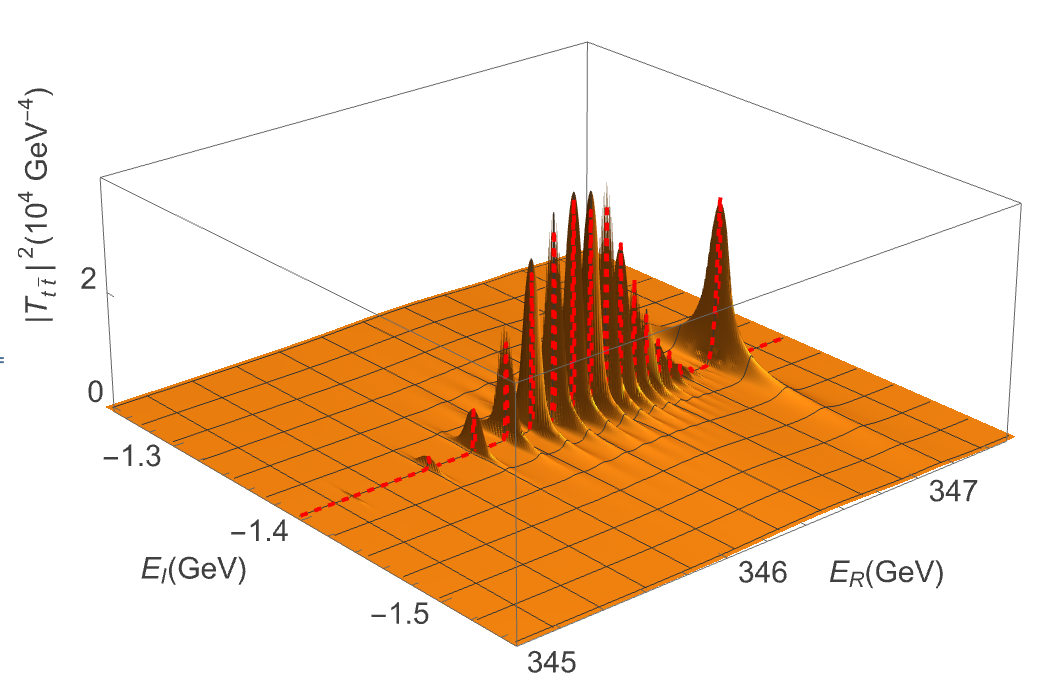}
    \caption{$S$-wave $T$-matrix in the complex energy plane in our baseline scenario with a top-quark width of 1.4\,GeV.}
    \label{fig:T-Z-14}
\end{figure}

{\it Complex-Pole Analysis.--}
The formal solution of the $T$-matrix equation can be written in operator form as
\be
\hat{T}(E) = \frac{{\hat{V}}}{1- G_2(E) \hat{V}}  \ , 
\ee
which highlights that the entire energy dependence is encoded in the uncorrelated two-body propagator, Eq.~\ref{GQQ}. In the small-width limit one has $G_{\ttb}(E,k)= (E-2\varepsilon_k\pm i\eta)^{-1}$ with infinitesimal $\eta$.
With an (approximately) energy independent top quark width one can carry out the integration over the relative-energy variable in Eq.~(\ref{GQQ}) to obtain 
\be
G_{\ttb}(E,k,\Gamma_t) = \frac{1}{E-2\varepsilon_t(k) +i \Gamma_t} \ .
\ee
This form enables a straightforward continuation to complex energies by replacing $E \to Z\equiv E_R +iE_I$. The resulting $S$-wave $T$-matrix in the complex energy plane for $\Gamma_t$=1.4\,GeV is shown in Fig.~\ref{fig:T-Z-14}. One finds that deep into the complex plane, at $E_I = \Gamma_t$\footnote{note that the definition of $E_I$ essentially corresponds to half the bound-state width, and thus $E_I=\Gamma_{\ttb}/2 = \Gamma_t$}, there still exists a well defined bound-state spectrum in terms of the $T$-matrix poles whose real parts
essentially agree with the masses on the real axis in the small-width limit. 
%


%
%

{\it Conclusions.--}
We have investigated the toponium spectrum using the $T$-matrix formalism with a Cornell potential benchmarked in previous applications to charmonia and bottomonia. Focusing on the $S$-wave channel, and utilizing the small-width limit for top quarks, the solutions of the Lippmann-Schwinger equation in momentum space exhibit an $S$-wave spectrum of about 20 (spin-degenerate) bound states whose spacing transits from Coulomb-like to equidistant string-like. For top-quark widths above $\sim$0.1 GeV, the spectral shape melts into a broad peak structure, whose location is significantly higher in mass than for the would-be the ground-state. This peak is essentially generated by the overlapping (broad) bound states close to the nominal threshold which are generated by the nonperturbative string interaction, and thus its strength is quite sensitive to the value of the string tension. It is also somewhat sensitive to the top quark width.
We have then carried out a complex-energy analysis for a realistic top width of 1.4\,GeV. It revealed that the pole structure from the small-width limit survives deep into the complex plane. This opens the possibility to relate the spectral properties on the real axis to the fundamental bound-state spectrum of toponium.

{\it Acknowledgments.--}
This work has been supported by the U.S. National Science Foundation under grant no. PHY-2514775 and the U.S. Department of Energy, Office of Science, Office of Nuclear Physics through the Topical Collaboration in Nuclear Theory on \textit{Heavy-Flavor Theory (HEFTY) for QCD Matter} under award no.~DE-SC0023547.


\bibliography{bibliography}

\end{document}